\documentclass[modern, twocolumn, trackchanges,export]{aastex631}
\let\splitbox\relax

\usepackage{adjustbox}

\usepackage[section]{placeins}
\usepackage{appendix}

\usepackage{float}
\usepackage{wrapfig}
\usepackage{array}

\begin{document}

\title{Hidden in Plain Sight: Searching for Globular Clusters Within JWST Observations of the PLCK G165.7+67.0 Galaxy Cluster}

\correspondingauthor{Tyler R. Hinrichs}
\email{trhinric@asu.edu}

\author[0009-0008-0376-3771]{Tyler R. Hinrichs}
\affiliation{School of Earth and Space Exploration, Arizona State University, Tempe, AZ 85287-6004, USA
}

\author[0000-0001-9394-6732]{Patrick S. Kamieneski}
\affiliation{School of Earth and Space Exploration, Arizona State University, Tempe, AZ 85287-6004, USA
}

\author[0000-0001-8156-6281]{Rogier A. Windhorst}
\affiliation{School of Earth and Space Exploration, Arizona State University, Tempe, AZ 85287-6004, USA
}

\author[0000-0003-3329-1337]{Seth H. Cohen}
\affiliation{School of Earth and Space Exploration, Arizona State University, Tempe, AZ 85287-6004, USA
}

\author[0000-0003-1625-8009]{Brenda L. Frye}
\affiliation{Steward Observatory, University of Arizona, 933 N Cherry Ave, Tucson, AZ, 85721-0009}

\author[0000-0001-6650-2853]{Timothy Carleton}
\affiliation{School of Earth and Space Exploration, Arizona State University, Tempe, AZ 85287-6004, USA
}

\author[0000-0002-2282-8795]{Massimo Pascale}
\affiliation{Department of Astronomy, University of California, 501 Campbell Hall \#3411, Berkeley, CA 94720, USA}

\author[0000-0001-9065-3926]{Jose M. Diego}
\affiliation{Instituto de F\'isica de Cantabria (CSIC-UC). Avenida. Los Castros
s/n. 39005 Santander, Spain}

\author[0000-0003-1268-5230]{Rolf A. Jansen}
\affiliation{School of Earth and Space Exploration, Arizona State University, Tempe, AZ 85287-6004, USA
}

\author[0000-0001-6265-0541]{Jessica Berkheimer}
\affiliation{School of Earth and Space Exploration, Arizona State University, Tempe, AZ 85287-6004, USA
}


\author[0000-0003-4875-6272]{Nathan J. Adams}
\affiliation{Jodrell Bank Centre for Astrophysics, Alan Turing Building, University of Manchester, Oxford Road, Manchester M13 9PL, UK}

\author[0000-0003-1949-7638]{Christopher J. Conselice} 
\affiliation{Jodrell Bank Centre for Astrophysics, Alan Turing Building,
University of Manchester, Oxford Road, Manchester M13 9PL, UK}

\author[0000-0001-9491-7327]{Simon P. Driver} 
\affiliation{International Centre for Radio Astronomy Research (ICRAR) and the
International Space Centre (ISC), The University of Western Australia, M468,
35 Stirling Highway, Crawley, WA 6009, Australia}

\author[0000-0002-7460-8460]{Nicholas Foo}
\affiliation{School of Earth and Space Exploration, Arizona State University, Tempe, AZ 85287-6004, USA
}

\author[0000-0003-3418-2482]{Nikhil Garuda}
\affiliation{Steward Observatory, University of Arizona, 933 N Cherry Ave, Tucson, AZ, 85721-0009}

\author[0000-0001-6145-5090]{Nimish P. Hathi}
\affiliation{Space Telescope Science Institute, 3700 San Martin Dr., Baltimore, MD 21218, USA}

\author[0000-0002-9984-4937]{Rachel Honor}
\affiliation{School of Earth and Space Exploration, Arizona State University, Tempe, AZ 85287-6004, USA
}

\author[0000-0002-6610-2048]{Anton M. Koekemoer}
\affiliation{Space Telescope Science Institute, 3700 San Martin Dr., Baltimore, MD 21218, USA}

\author[0000-0002-6150-833X]{Rafael Ortiz III}
\affiliation{School of Earth and Space Exploration, Arizona State University, Tempe, AZ 85287-6004, USA
}

\author[0000-0002-8556-4280]{Marta Reina-Campos}
\affiliation{Canadian Institute for Theoretical Astrophysics (CITA), University of Toronto, 60 St George St, Toronto, M5S 3H8, Canada}
\affiliation{Department of Physics \& Astronomy, McMaster University, 1280 Main Street West, Hamilton, L8S 4M1, Canada}

\author[0000-0003-0429-3579]{Aaron S. G. Robotham}
\affiliation{International Centre for Radio Astronomy Research (ICRAR) and the International Space Centre (ISC), The University of Western Australia, M468, 35 Stirling Highway, Crawley, WA 6009, Australia}

\author[0000-0002-7265-7920]{Jake S. Summers}
\affiliation{School of Earth and Space Exploration, Arizona State University, Tempe, AZ 85287-6004, USA
}

\author[0000-0001-7592-7714]{Haojing Yan}
\affiliation{Department of Physics and Astronomy, University of Missouri, Columbia, MO 65211, USA}

\author[0000-0001-8762-5772]{William E. Harris}
\affiliation{Department of Physics \& Astronomy, McMaster University, 1280 Main Street West, Hamilton, L8S 4M1, Canada}

\begin{abstract}
Although the James Webb Space Telescope (JWST) has received much attention for its ability to search deeper into the cosmos than ever before, it also enhances our capability to study objects closer to us in the Universe. We apply a methodology of subtracting intracluster light to the PLCK G165.7+67.0 (G165; $z$ = 0.35) cluster, revealing a population of unresolved point-like sources including globular clusters (GCs). By applying a fitting algorithm in color space used to select galaxy cluster members, we uncover over 900 globular cluster candidates from our point source sample.
We also identify candidates by estimating the contribution of interlopers to the point source sample, yielding an estimate of 
793$\pm$ 83
globular cluster candidates. 
We find the color-selected sources to be approximately correlated spatially with the intracluster light and lensing mass of the cluster. The observed luminosity function of the sources shows a turnover point fainter than the completeness limit, so we use fixed-parameter curve fitting models to predict a K-corrected turnover point between $-9.4 \leq M_{\rm F200W} \leq -10.7$ mag, although we predict the expected K-corrected turnover point should be closer to $-7.7 \leq M_{\rm F200W} \leq -8.4$ mag. We discuss the dynamical state of this disturbed galaxy cluster with a bimodal mass distribution using the spatial distribution of GC candidates and find that the radial profiles of our color-selected GC candidates are very consistent with the lensing-derived surface mass density at $>$50 kpc.

\end{abstract}

\keywords{Galaxy clusters(584) --- Globular star clusters(656) --- James Webb Space Telescope(2291)}

\section{Introduction} \label{sec:intro}

PLCK G165.7+67.0 (hereafter G165) is a massive galaxy cluster at $z$ = 0.35 that was discovered with {\it Planck} as a bright submillimeter point source, resulting from a highly-magnified dusty star-forming galaxy behind the cluster at $z = 2.24$ \citep{Canameras2015, Harrington2016, Planck2016}. It is notable for a clearly bimodal spatial distribution of cluster galaxies, which results in a region of high magnification between the two primary mass components. G165 has one of the apparently brightest dusty star forming galaxies on the sky, among other prominent giant arcs and multiply-imaged galaxies \citep{Canameras2018, Frye2019, Pascale2022a, Frye2024:aa, Kamieneski2024}. One of these arcs contains a triply-imaged type Ia supernova at $z=1.78$, SN H0pe \citep{Frye2023, Frye2024:aa}, which has recently been used to derive Hubble's constant \citep{Pascale2025}.

In the era of JWST, studies have revealed new insights into globular cluster (GC) systems in galaxy cluster environments such as SMACS 0723.3-7327, Abell 2744, El Gordo, MACS J0417.5-1154, and MACS J0416.1-2403 \citep{Lee2022, Faisst2022, Pascale2022b, Harris2023p1, Diego2024, Martis2024:aa, Harris2024p2, Berkheimer2025, Harris2025a, Harris2025b}. Within G165, there is a clear population of point-like sources embedded in the most luminous cluster galaxies. At a distance of G165, these point sources are unresolved and include both GCs and Ultra Compact Dwarfs (UCDs) given their small physical sizes \citep{Larsen2001}. Deep observations of G165 from the PEARLS (Prime Extragalactic Areas for Reionization and Lensing Science, \citealt{Windhorst2023:aa}) GTO program and in two additional epochs of JWST observations \citep{Frye2024:aa}, with a total exposure time of 5,626 seconds in F200W, allow for excellent sensitivity and the ability to subtract intracluster light \citep{Pascale2022a} for the study of point sources buried within.

Intracluster light (ICL) is the diffuse light concentrated within clusters of galaxies, created by stars bound to the gravitational potential of the galaxy cluster \citep{Montes2022}. Both ICL and GCs result from a common origin as stars are stripped away from their parent halos \citep{Diego2023}. Given this, the spatial distribution of GCs embedded within galaxy clusters can be compared with the distribution of the ICL, as independent probes of the total mass distribution. Likewise, the distribution of the ICL and GCs can further be compared to lens models, predicting the amount of dark matter present within the galaxy cluster, as done with SMACS 0723.3-7327 (hereafter SMACS0723) and MACS J0416.1-2403 (hereafter MACS0416) \citep{Pascale2022b,Diego2023, Diego2024}. However, as found recently through simulations by \citet{Butler2025}, the ICL can be a complex and biased tracer of the underlying dark matter distribution, with intracluster stars probing a more centrally-concentrated density profile than the dark matter halo.

The globular cluster luminosity function (GCLF) is the distribution of magnitudes of GCs with a peak \citep{Harris1991, Rejkuba2012}, referred to as the turnover point. In nearby galaxies, the turnover point is approximately $M_{V} \simeq -$7.5 mag \citep{Kundu2001, Harris2009b, Rejkuba2012} in the V band, making the GCLF a well-established distance indicator, given a  precision of 0.1$-$0.2 mag \citep{Ostriker1997,Mei2005, Villegas2010,Rejkuba2012}. Studies have shown that there is a shallow correlation between the turnover point and host galaxy luminosity, which is relevant for bright cluster galaxies \citep{Villegas2010, Harris2014}. 
However, the lookback time cancels out this effect, making $M_{V} \simeq -$7.5 mag still an accurate estimator of the characteristic GCLF peak. Relatively few studies have so far examined the GCLF in galaxy cluster fields down to the sensitivity regularly achieved with JWST NIRCam beyond the local Universe. In this work, we model the GCLF in a galaxy cluster environment and make comparison to other works at a similar redshift, in order to further examine how consistent the GCLF is in the dense central regions of galaxy clusters.

The paper is organized as follows: Section \ref{sec:overview} describes the data and technique used in selecting our sample of GCs. Section \ref{sec:Results} discusses the results from the selected GCs, including the GCLF and comparison of spatial distribution to the modeled light. Finally, Section \ref{sec:conc} summarizes our conclusions and makes predictions for future progress that can be made in understanding globular clusters in lensing galaxy cluster fields. All images in this work are north-aligned. Unless otherwise noted, we use the AB system for our magnitudes \citep{Oke1983}. We assume a flat cosmological model with $H_0$ = 67.4 km s$^{-1}$ Mpc$^{-1}$, $\Omega_M$ = 0.315 and $\Omega_\Lambda$ = 0.685 \citep{Planck2020}, giving a luminosity distance $d_L$ = 1921.3 Mpc, angular distance $d_A$ = 1054.2 Mpc, and 0.195$\arcsec$ per kpc at $z$ = 0.35 \citep{Wright2006}.

\section{Data and Photometry} \label{sec:overview}
\subsection{Data} \label{subsec:data}
In this study, we use JWST NIRCam imaging of G165. In summary, G165 was observed in 3 epochs in 2023. The first epoch (8 NIRCam filters) was observed as a part of the PEARLS program (PID 1176; PI: R. Windhorst) on March 30, 2023. The second and third epochs were observed as a part of a JWST DDT Program (PID 4446; PI: B. Frye) on April 22, 2023 (6 NIRCam filters) and May 9, 2023 (6 NIRCam filters), respectively. A further detailed description of this data is given by \citet{Windhorst2023:aa} and \citet{Frye2024:aa}. We use images stacked with all 3 epochs in our analysis, allowing us to detect fainter sources than we would by using only a single epoch. In this study, we use these 6 commonly used JWST NIRCam filters in the observations of G165: F090W, F150W, F200W, F277W, F356W, F444W (see Table 1 in \cite{Frye2024:aa} for exposure times). The observations of G165 have been processed by the PEARLS team \citep{Windhorst2023:aa} using the JWST STScI pipeline, with additional attention to removal of 1/f noise and ``wisps`` \citep{Robotham2023}. All images used are also drizzled onto a 30 mas resolution. For our selection of GC candidates and redshift fitting, we use images that are PSF-convolved to the filter with the coarsest resolution, F444W, using the process described in \cite{Frye2024:aa} (see also \citealt{Pascale2022a}). For all other portions of the paper, including our point source identification, completeness limit tests, and GCLF analysis, we use the original images that have native PSFs.

\subsection{ICL Subtraction using ProPane} \label{subsec:ICLsub}
Subtracting the bright, smooth ICL component of our imaging can improve our ability to detect the faint sources closer to cluster cores. Previously, ICL subtraction of G165 was done in \citet{Pascale2022a} using \textsc{Galfit} \citep{GALFIT}. Although this method performs well in subtracting the light from galaxy cluster members, negative flux regions are left behind (even after residual corrections), interfering with our ability to detect point sources close to the cluster members. 

\begin{figure*}[ht!]
\includegraphics[width=1\textwidth]{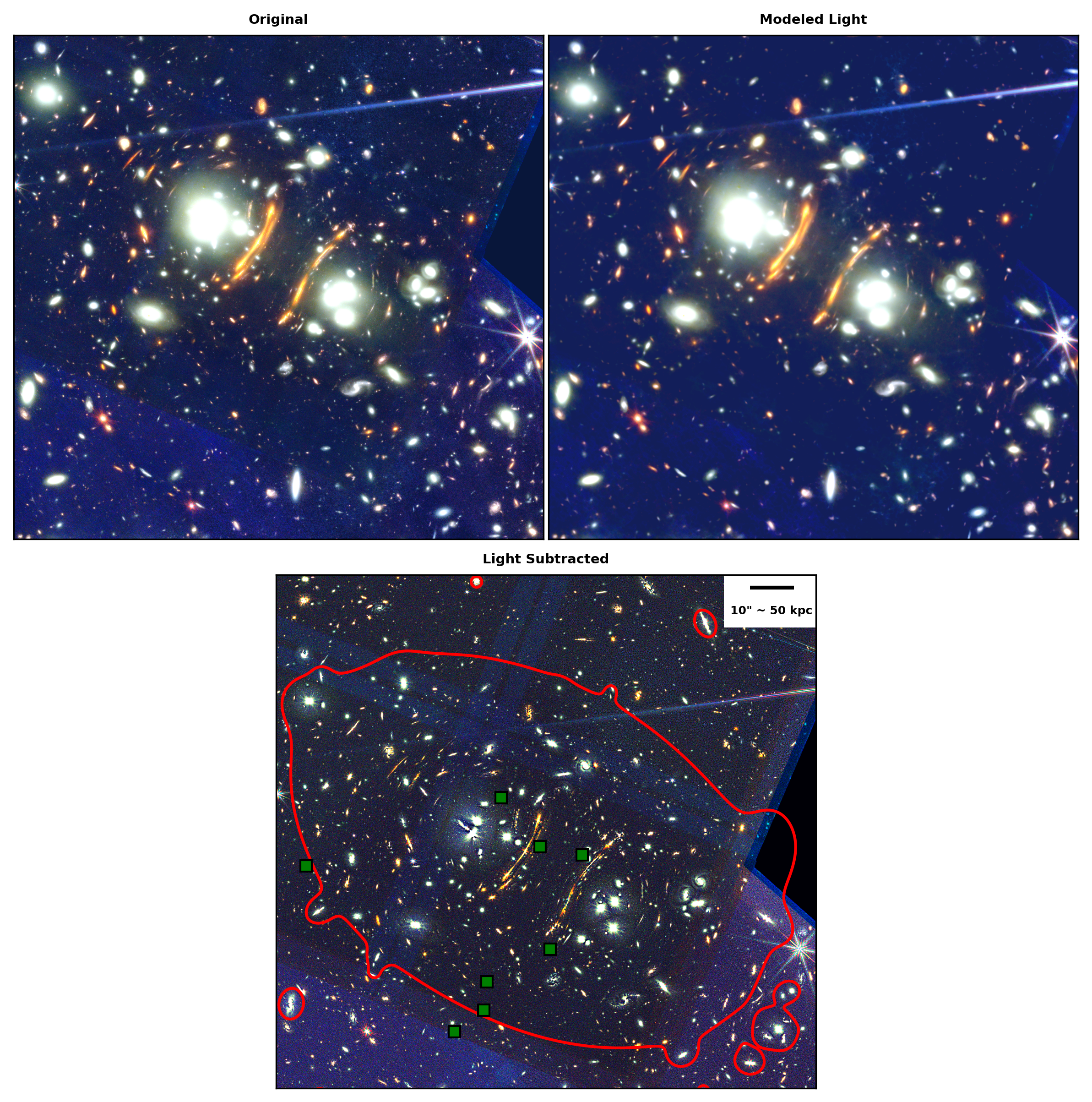}
\centering
\caption{
RGB image cutouts of the G165 cluster throughout the ICL subtraction process. All six filters are used with (R,G,B) = (F356W + F444W,  F200W + F277W, F090W + F150W). \emph{Top Left Panel}: Original image before any subtraction. \emph{Top Right Panel}: Modeled light using ProPane. \emph{Bottom Panel}: Final light-subtracted image with a single red contour showing the region where the lens model-derived convergence is $\kappa = 0.25$ \citep{Kamieneski2024}. The green squares are the 13 objects with well-fit photometric redshifts between $0.3 < z_{phot} < 0.4$ and the visible blue border in the image is due to lower exposures and dithering effects in the longer wavelength filters.
\label{fig:images}
}
\end{figure*}

Other studies of GC systems in galaxy clusters (e.g., \citealt{Lee2022, Keel2023:aa, Martis2024:aa}) use different ICL subtraction techniques, including isophote modeling \citep{Isophote1987:aa} and median ring filtering \citep{Secker1995}, to model and subtract the light around galaxies. As described in Appendix \ref{app:Isophote}, these methods applied to G165 result in significant amounts of oversubtraction, especially between cluster members, interfering with the ability to detect faint GCs within the field. To mitigate this issue, we use \textsc{ProPane} \citep{Robotham2024:aa} to model and subtract the diffuse extended light in G165. The propaneLocalMed function in \textsc{ProPane} takes a median of a defined grid size around each pixel, modeling the ICL and bright galactic light of the field. Through testing of bigger and smaller grid sizes, we found that an 11 by 11 pixel grid size performed best in detecting point sources. Using this function with 5 iterations per image and constraining the maximum flux change per pixel between iterations to the RMS of the background (preventing oversubtraction), we modeled and subtracted this light in all images, including a separate treatment for unconvolved and convolved images. While this filter includes all diffuse light in the image, which might originate from sources in front of and behind the galaxy cluster, it is dominated by the ICL in the regions nearest to the most luminous cluster galaxies. The RGB color image cutouts (see Figure 1 of \citealt{Frye2024:aa} for full images) showing this process are shown in Figure \ref{fig:images}, created with color scaling according to the Trilogy\footnote{\href{https://www.stsci.edu/~dcoe/trilogy/Intro.html}{https://www.stsci.edu/$\sim$dcoe/trilogy/Intro.html}} prescription \citep{Coe2024:aa}. As seen by this color image, there is still significant residual light near the central cores of the bright galaxy cluster members, extending out to around 10 kpc from the cores, an implication which is considered throughout the study. 

\subsection{Point Source Identification} \label{subsec:photometry}
Using the PSF-convolved ICL-subtracted images, we run \textsc{Source Extractor} (\textsc{SExtractor}, \citealt{Bertin1996:aa}) in dual mode for each filter, using the ICL-subtracted F200W image as the detection-image, given that GCs are brightest in this filter at the redshift of G165. To capture the faintest sources in F200W, we perform $0\farcs18$ 
(6 pixels) diameter aperture photometry on the detection-image using SNR $>$ 2 per pixel and only include sources with a minimum of 4 connected pixels. Likewise, we perform $0\farcs15$ and $0\farcs3$ diameter aperture photometry on the detection-image, used to measure object concentrations to identify point sources.
Kron aperture photometry (MAG$\_$AUTO) was also obtained on each of the PSF convolved filters to be used for our color-color diagrams. We find negligible differences in the color-color diagrams using Kron apertures compared to $0\farcs18$ diameter apertures (specifically, sigma-clipped (3$\sigma$) median differences of 0.03 mag in F150W$-$F200W and 0.01 mag in F277W$-$F356W). Our ICL subtraction method, along with 3 epochs' worth of JWST data, allow us to detect objects with magnitudes closer to $m_{\rm F200W} \sim 30$ mag (AUTO) compared to $m_{\rm F160W} \sim 27$ mag (AUTO) using HST \citep{Pascale2022a}.

\begin{figure*}[ht!]
\centering
\includegraphics[width=1\textwidth]{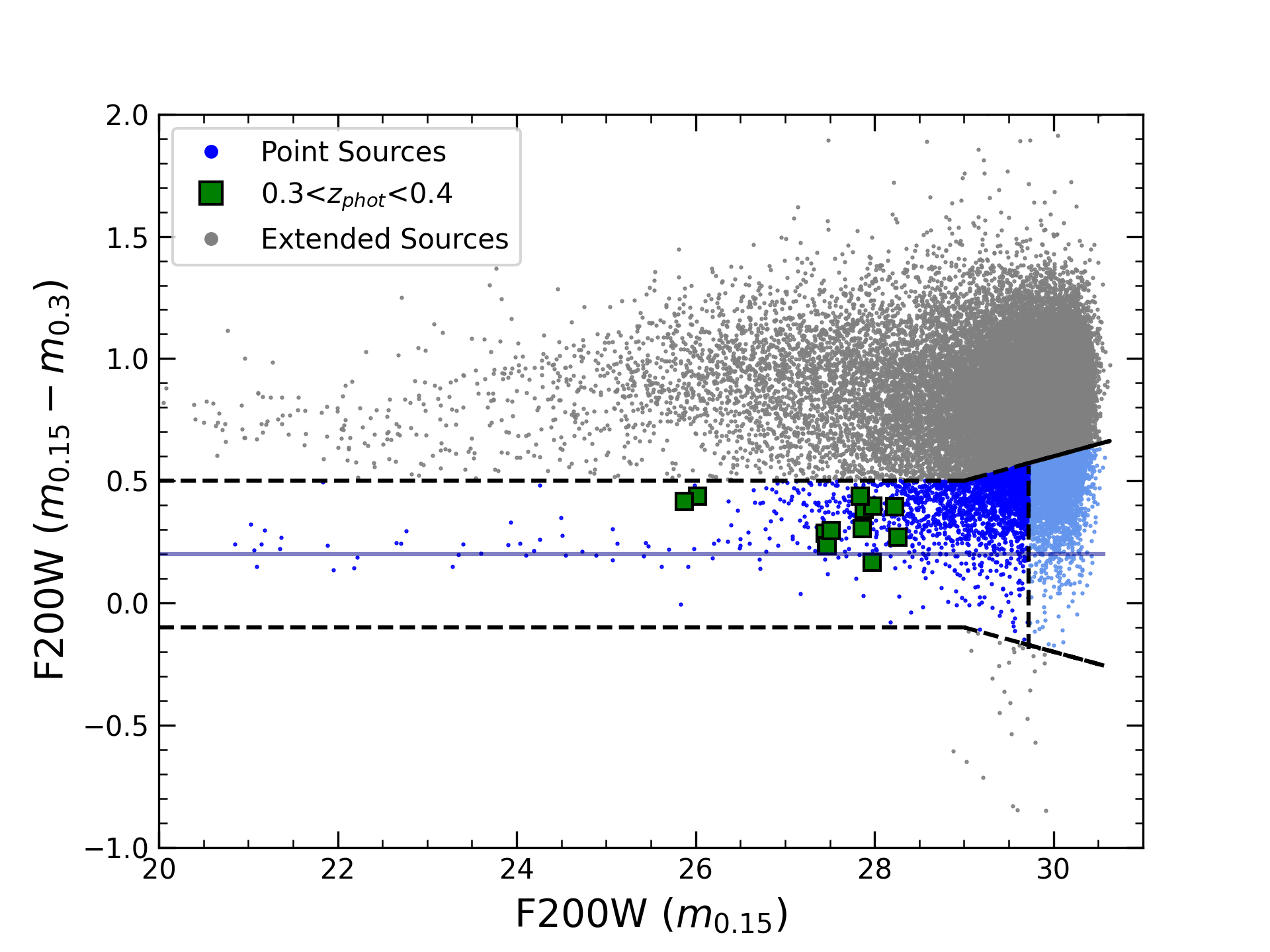}
\caption{
Compactness selection of point sources using the photometric difference of aperture diameters $0\farcs15$ and $0\farcs3$ in the F200W filter as a function of F200W magnitude in $0\farcs15$ diameter aperture. Our selection in dark blue includes all point sources in the field, including GCs, interloper UCDs, and foreground stars. Within this selection, we include the 13 objects (green squares) with well-fit photometric redshifts between $0.3 < z_{phot} < 0.4$. The vertical black line at 29.73 mag indicates the 50$\%$ completeness limit for $0\farcs18$ aperture diameters, and any point sources fainter than this limit are still used, but highly uncertain. The gray points are removed objects that do not meet the compactness threshold (horizontal dashed black lines centered around the solid line).
\label{fig:compactness}
}
\end{figure*}

Globular clusters have half-light radii of $\sim2.5{\rm ~pc}$ \citep{Harris2010}, so they appear as unresolved point sources at the distance of G165. To identify these point sources in our catalog, we make a selection based on the compactness of sources in F200W. The compactness of the sources can be defined by the difference of fluxes between two aperture sizes, otherwise known as the concentration index \citep{Peng2011}. Throughout the past decade, the concentration index has been used in the selection of GC candidates in many ground-based and space-based studies of galaxy clusters \citep{Durrell2014, Lim2018,LaMarca2022,Toloba2023, Martis2024:aa, Saifollahi2025, Kluge2025, Marleau2025}. Following \citet{Mirabile2024} and \citet{Saifollahi2025}, we make a point source selection with magnitude-dependent broadening using the difference in brightness between $0\farcs15$ and $0\farcs30$ aperture diameters for the F200W filter, as shown in Figure \ref{fig:compactness}.

Our compactness selection is based on the clearer dichotomy in point sources that exists brighter than $m_{0.15}\sim$27.5 mag, which becomes less apparent at fainter magnitudes. Between 20 $\leq$ $m_{0.15}$ $\leq$ 29 mag, we select point sources that have a compactness of -0.1 $\leq$ $m_{0.15}-m_{0.3}$ $\leq$ 0.5, centered on the horizontal sequence at $m_{0.15}-m_{0.3}$ $\sim$ 0.2, shown by the solid line in Figure \ref{fig:compactness}. For $m_{0.15}$ $\geq$ 29 mag, we select point sources with a compactness of $-0.1 \times$$m_{0.15}$+2.8 $\leq$ $m_{0.15}-m_{0.3}$ $\leq$ $0.1 \times$$m_{0.15}$-2.4. Using this magnitude-dependent broadening allows us to include more point sources at faint magnitudes where the recovery rate of point sources starts to decrease. This selection also encapsulates the median $m_{0.15}-m_{0.3}$ values of the recovered stars from our artificial star test in Section \ref{subsec: completeness} up to the 50$\%$ completeness limit at 29.73 mag, although this limit is found using slightly larger aperture diameter sizes ($0\farcs18$). Any point sources past this limit (shown by the light blue points in Figure \ref{fig:compactness}) are still used as needed in the analysis, but are more uncertain and are heavily downweighted by our completeness function. This selection yields a sample of 6413 point sources in the field, filtered from the 47321 detected sources from the initial photometry.

Furthermore, \citet{Harris2023p1} classifies UCDs as point sources brighter than M$_{\rm F150W} = -13.3$ mag in Abell 2744. Following \citet{Berkheimer2024}, this converts to M$_{\rm F200W} = -13.3 -0.32 = -13.62$ mag (m$_{\rm F200W} = 27.56$ mag) in G165, where 0.32 is the mean F150W-F200W of our point source sample. From our point source selection, it is clear that the majority of our sample are possible GC candidates, with only $<$2$\%$ being potential UCDs according to this definition.

\subsection{Photometric Redshift Analysis} \label{subsec:EAZY}
\begin{figure*}[ht!]
\centering
\includegraphics[width=1\textwidth]{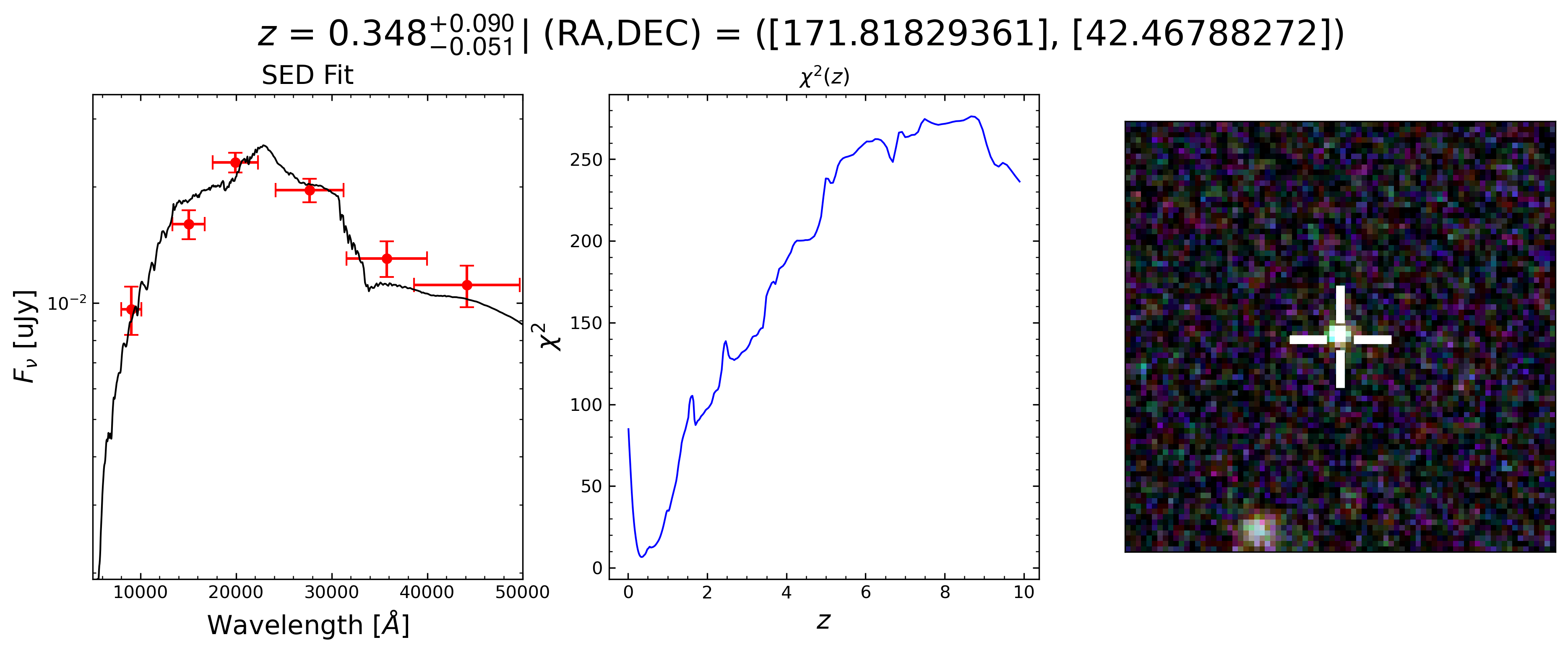}
\caption{
Example of SED fitting of object close to the cluster redshift, ran by EAZY \citep{BrammerEAZY}. The title shows the redshift found at the minimum $\chi^2$ and the RA/DEC if the point source. The computed photometric redshift of this particular source is at $z$ = 0.348. \emph{Left Panel}: SED fit using $F_{\nu}$ [$\mu$Jy] vs filter wavelength [\AA]. \emph{Middle Panel}: $\chi^2$ values at different redshifts, used to predict the best-fit redshift for the source. \emph{Right Panel}: RGB image of object, shown within the crosshair at the center of the image.
\label{fig:SEDFIT}
}
\end{figure*}

One method of further identifying only the GCs in our sample is to estimate the redshifts of our point sources using \textsc{EAZY} \citep{BrammerEAZY}. When fitting our point sources, we find the majority cannot be adequately fit by the templates in \textsc{EAZY}, due to their faint magnitudes ($m_{\rm F200W} > 28$ mag). Specifically, we use the CWW+KIN \citep{CWW, KIN} template implemented in \textsc{EAZY} using a single template fit. We do find 13 objects that have a fit photometric redshift between 0.3 $<z_{phot}<$ 0.4 and 68th percentile confidence intervals bounded on either side by 0.25 and 0.45, indicating they have a photometric redshift solution very close to the cluster redshift. We include these objects in Figures \ref{fig:images} and \ref{fig:compactness}, showing that they are point sources and most are located spatially in the vicinity of the brightest cluster members. Instead of using \textsc{EAZY} to select GC candidates, we focus on the brighter point sources that are fit at the cluster redshift in order to assess where they lie in color space around the expected 1.6 $\mu$m bump.
The 1.6 $\mu$m bump is caused by the low absorption of H$^-$ in the atmosphere of stars at this wavelength, which has been shown to be an accurate photometric redshift indicator \citep{Simpson1999, Sawicki2002, Frye2024:aa}. Figure \ref{fig:SEDFIT} shows the redshifted 1.6 $\mu$m bump between F200W and F277W, indicating that F150W, F200W, F277W, and F356W would offer an efficient color selection of GCs at the redshift of G165.

\begin{figure*}[th!]
\centering
\includegraphics[width=0.8\textwidth]{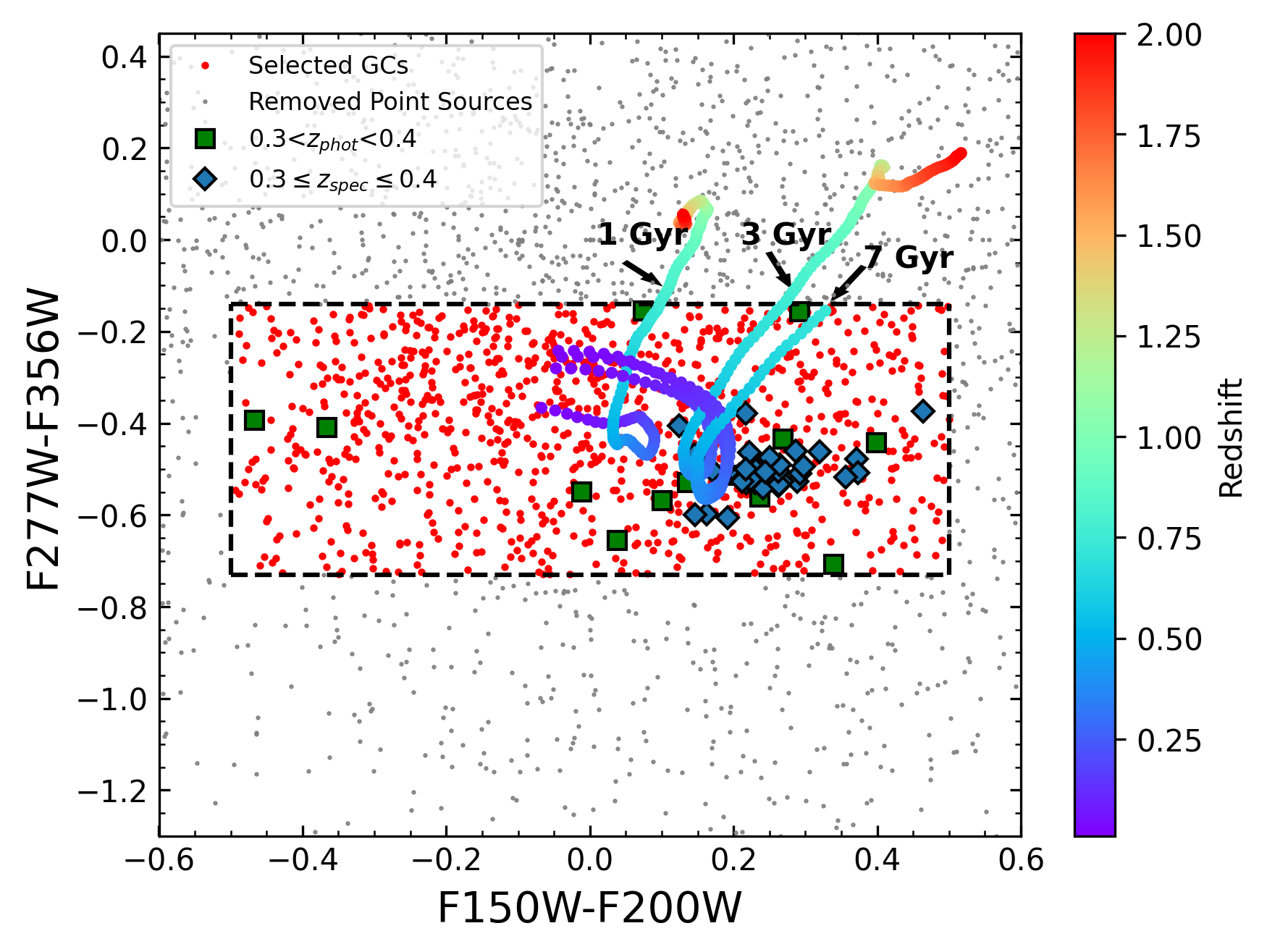}
\caption{
 Color-color diagram using F277W-F356W vs F150W-F200W to select G165 cluster members (within the compactness criterion) at $z$ = 0.35. Gray points show the point sources that do not pass our color selection, red points within the black dashed box show the selected globular cluster candidates determined to be likely at the redshift of the galaxy cluster, green squares are point sources with photometric redshifts between 0.3 and 0.4, and blue diamonds are objects with spectroscopic redshifts between 0.3 and 0.4 (centered on the cluster redshift). BC03 stellar populations \citep{BruzualCharlot2003} at 1, 3, and 7 Gyr are shown by the rainbow-colored points, ranging from redshifts of 0 to 2 encoded by the colorbar.
\label{fig:GC Selection}
}
\end{figure*}

\subsection{Globular Cluster Selection} \label{subsec:Selection}
\subsubsection{Color Selection}
\label{subsubsec: Color Selection}
For a more robust and complete selection of GCs within the G165 cluster, we adopt the method of selecting galaxy cluster members using a color-color diagram along with the colors of stellar populations at different ages \citep{Frye2024:aa}. We use Kron apertures in F150W, F200W, F277W, and F356W in order to capture the 1.6 $\mu$m bump, as seen in Figure \ref{fig:SEDFIT}. In Figure \ref{fig:GC Selection}, we show the redshift tracks for 1, 3, and 7 Gyr old stellar populations (BC03, \citealt{BruzualCharlot2003}) in color space, which assume solar metallicity. In addition to the stellar populations, we include the colors of the 13 point sources that have photometric redshifts between 0.3 $<z_{phot}<$ 0.4. Guided by these photometric redshifts and stellar population tracks, we select all sources within a threshold of $-0.5 \le$ F150W-F200W $\le 0.5$ and $-0.73 \le$ F277W-F356W $\le -0.14$, yielding 972 GC candidates. The motivation behind these selection bounds and basic properties of the point sources with well-fit photometric redshifts is further described in Appendix \ref{app:Bounds}. To complement our selection, we show the colors of 39 galaxies in the cluster field with spectroscopic redshifts within $0.3<z<0.4$, found using NIRSpec, further described in \citealt{Frye2024:aa}. This selection is shown by the red points within the black dashed box in Figure \ref{fig:GC Selection}.

\subsubsection{Subtraction of Interlopers to Point Source Catalog}
We also estimate the abundance of GC candidates in G165 by subtracting the magnitude densities of ``background" point sources from ``cluster" point sources, a method that has been widely used to determine GC population sizes \citep{Secker1993,Gomez2001,Peng2011,Hudson2018, Lim2018, Amorisco2018}. Cluster point sources are identified as being within the region where $\kappa > 0.25$, found from a recent lensing model for G165 \citep{Kamieneski2024}. Likewise, background point sources are identified in the outer regions, where $\kappa < 0.25$. This threshold is chosen based on the locations of the 21 families of lensed images that constrain the lens model \citep{Frye2024:aa}. For the \citet{Kamieneski2024} model, $\kappa \approx 0.25$ corresponds to a projected distance from the two cluster halos that is at least twice that of the furthest multiply-imaged arc, and therefore far from the known strong lensing regime. This region is shown by the single red contour in Figure \ref{fig:images}. 
It is still feasible that even these lower-convergence regions may contain a non-insignificant portion of the cluster mass, meaning that the designated ``background" point sources may include some globular clusters that pertain to the lensing cluster. However, the positioning of the cluster center within the small field of view of NIRCam (relative to the $\gtrsim 1$ Mpc virial radius of galaxy clusters of a similar mass) is not ideal and does not offer a true background. As we discuss in Section \ref{subsub:numberGCs}, the low number of recovered GCs relative to what is expected for G165 may be a result of interloper over-subtraction.
After identifying each group of point sources, we divide the number counts per 0.2 magnitude by their respective areas to obtain number counts per unit area and subtract the background number counts from the cluster-centered counterpart in order to isolate only point sources that pertain to the cluster. Finally, we multiply these counts per unit area by the cluster area (where $\kappa > 0.25$) to compare evenly with our color-selected GC candidate sample. This process of estimating magnitude densities of cluster GC candidates in each 0.2 magnitude bin is described by

\begin{equation}
    N_{ang} = (\frac{N_{cluster}}{A_{cluster}} -  \frac{N_{background}}{A_{background}}) \cdot A_{cluster}
 \label{Equation1},
\end{equation}
\noindent
where $N_{ang}$ is the total number of estimated GC candidates per 0.2 magnitude bin, $N_{cluster}$ is the number of point sources per 0.2 magnitude bin within the region where $\kappa > 0.25$ with area $A_{cluster}$, and $N_{background}$ is the number of point sources in each 0.2 magnitude bin within the region where $\kappa < 0.25$ with area $A_{background}$. For G165, $A_{cluster} = 1.75$ [arcmin$^2$] and $A_{background} = 8.76$ [arcmin$^2$], giving a total field area of 10.51 [arcmin$^2$]. This alternative estimation yields a total of 793 GC candidates.

\section{Results} \label{sec:Results}
\subsection{Completeness Limit}
\label{subsec: completeness}

\begin{figure}[H]
\centering
\includegraphics[width=0.5\textwidth]{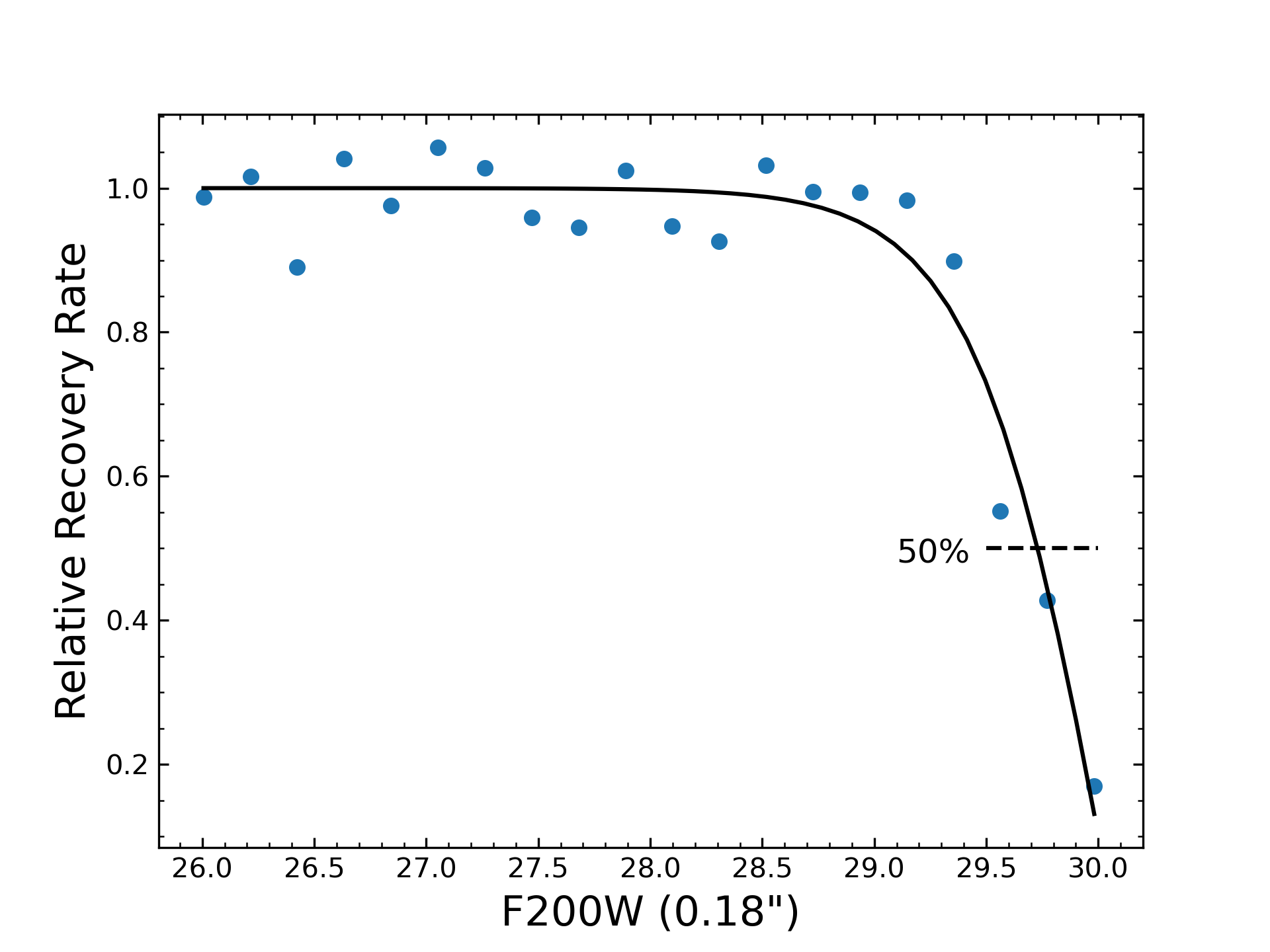}
\caption{
Recovery rate as a function of flux for injected artificial point sources. The blue points show the recovery rate of the artificial objects at different magnitudes. The solid black line shows the best-fit solution using Equation \ref{Equation2}, with parameters $\beta$ = 3.3 and $m_{1}$ = 30.06. We find that the completeness limit at 50 percent is $m_{\rm F200W}$ = 29.73 mag, indicated by the dashed line. Note, real sources mistaken for test sources leads to values $>$1.
\label{fig:Completeness Limit}
}
\end{figure}

We use artificial stars to derive the 50$\%$ detection completeness limit \citep{Harris2023p1}, using \textsc{SExtractor} and \textsc{Photutils} \citep{Bradley2016}. In this process, we inserted 2000 artificial stars into the ICL-subtracted F200W image over 5 iterations using \textsc{Photutils}. We use a constant luminosity function between 26 and 32 magnitude for the brightness of the stars, distribute them randomly throughout the field with a minimum separation of 6 pixels, and assume a Gaussian PSF with a standard deviation matching the F200W PSF. After insertion of the stars, source detection and $0\farcs18$ aperture diameter photometry was measured using the same parameters as for the GC candidates themselves, as described in Section \ref{subsec:photometry}. The recovery rate of these stars in each iteration was found by dividing the amount of stars recovered by the number of inserted stars in each 0.2 magnitude bin. The total recovery rate was then found by finding the average recovery rate in each 0.2 magnitude bin throughout the 5 iterations. The final recovery rate is accurately matched by a tangent hyperbolic function \citep{Harris2023p1},
\begin{equation}
\setlength\abovedisplayskip{3pt}
\setlength\belowdisplayskip{3pt}
f(m) = \frac{2}{1 + e^{\beta(m-m_{1})}} - 1
\label{Equation2},
\end{equation}
\noindent
with best-fit parameters of $\beta$ = 3.3 $\pm$ 0.4, $m_{1}$ = 30.06 $\pm$ 0.04 mag. Using these parameters, we find that the completeness limit at 50 percent is $m_{\rm F200W}$ = 29.73 mag (Figure \ref{fig:Completeness Limit}), compared to $m_{\rm F200W}$ = 28.54 mag found in \citet{Frye2024:aa} for G165, resulting from the differing approach and detection threshold (a 5$\sigma$ detection threshold is used in \citealt{Frye2024:aa}).
Furthermore, we do not implement a spatially-varying completeness function (e.g. \citealt{Harris2024p2, HarrisNeural}) for now, but this will be explored in future work using deeper data.

\begin{figure*}[htb!]
\centering
\includegraphics[width=0.5\textwidth]{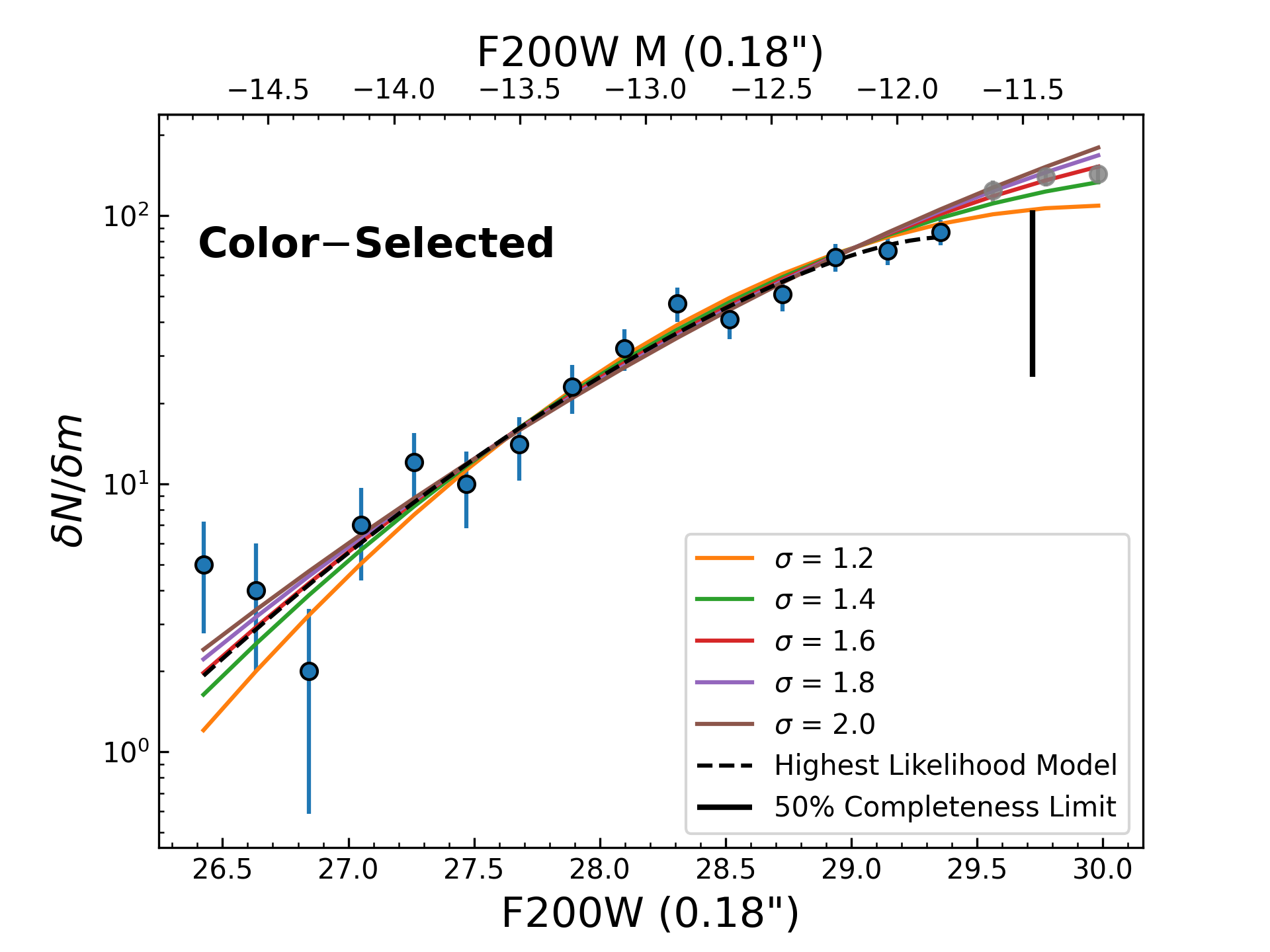}\hfill \includegraphics[width=0.5\textwidth]{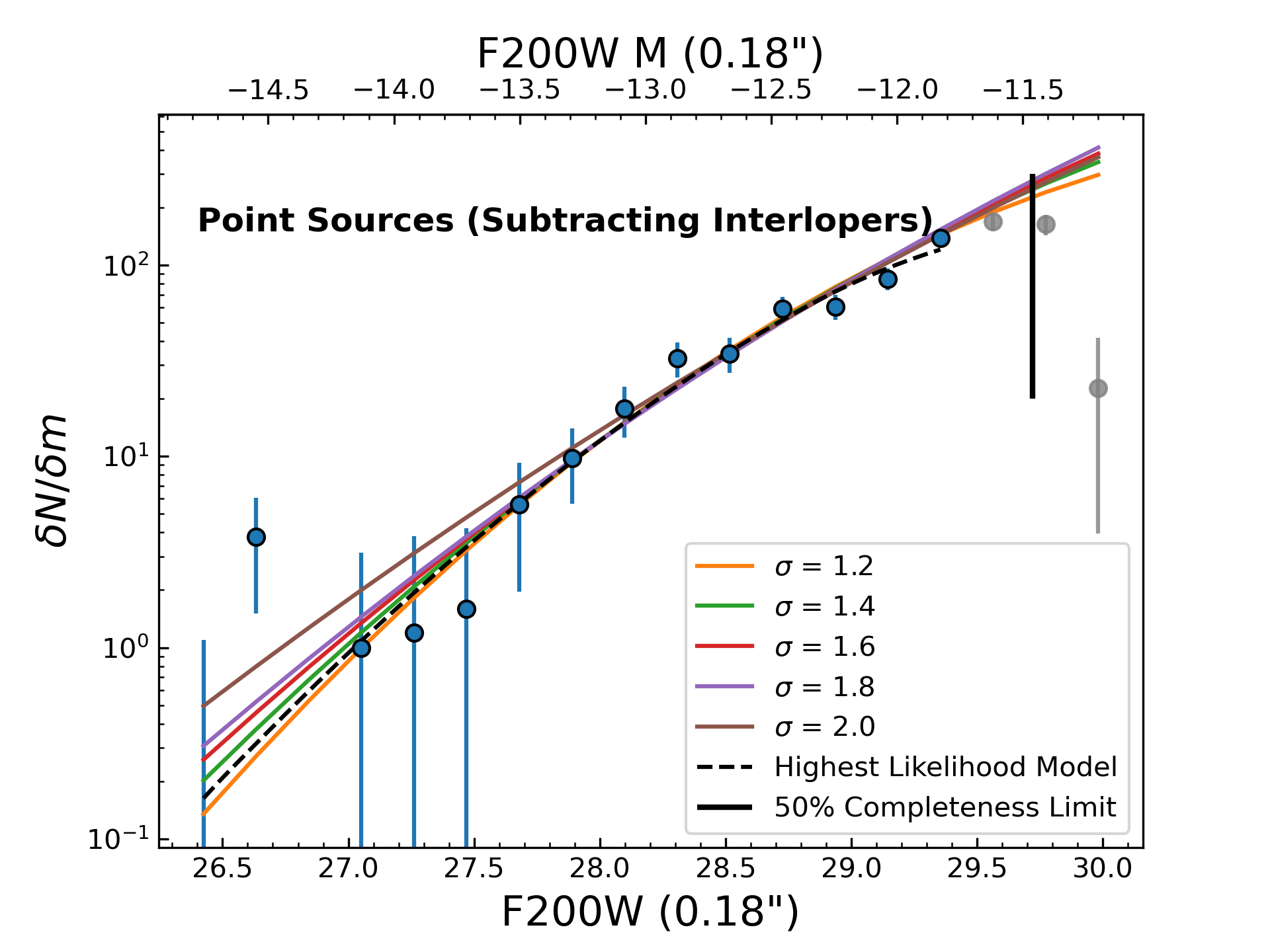}
\caption{
Luminosity functions of the GCs in G165, showing the GC candidates selected by color (\emph{left}) and GC candidates chosen by subtracting contaminating interloper objects from the sample of point sources (\emph{right}). The blue data points show the counts from 26.5 to 30.0 magnitude, with error bars showing Poisson uncertainties, and the gray data points are counts of magnitudes with $<80 \%$ completeness that are left out from the fits. The dashed black line shows the best-fit to the observed GCLF with $\sigma$ as a free parameter, showing that the empirical peak (i.e. intrinsic GCLF convolved with the recovery rate) is slightly brighter than the 50$\%$ completeness limit, indicated by the solid black line. The solid colored lines show the intrinsic GCLF of G165 at different fixed sigma values if recovery is not taken into account in the fit.
 \label{fig:GCLF}
}
\end{figure*}

\subsection{Luminosity Function of GCs} 
\label{subsec:GCLF}
Figure \ref{fig:GCLF} shows the GCLF of G165 using two different methods of selecting GCs. The left panel shows the GCLF for our color-selected GC candidates, whose selection is described in Section \ref{subsec:Selection}. The right panel shows the GCLF of GC candidates when subtracting interloper objects from our total point source sample, further described in Section \ref{subsub:numberGCs}.  Following \citet{Harris2023p1}, we fit the GCLFs using \textsc{emcee} \citep{emcee} with a modified hyperbolic tangent function,

\begin{equation}
 \frac{\delta N}{\delta m} = \frac{N}{\sqrt{2\pi}\sigma}e^\frac{-(m-m_0)^2}{2\sigma^2} \cdot f(m)
 \label{Equation3},
\end{equation}

where $N$ is the total number of globular clusters in the field, $\sigma$ is the dispersion of the GCLF, $m_{0}$ is the true turnover magnitude, and $f(m)$ is the recovery rate given by Equation \ref{Equation2} with fixed best-fit parameters derived by the artificial star recovery test. When the observed peak of the luminosity function does not quite reach the true turnover magnitude, as with our current observations of G165, $m_0$ and $\sigma$ are highly degenerate and uncertain \citep{Hanes1987}. To mitigate this, we make a prediction of the true $N$ and $m_{0}$ values for broad fixed $\sigma$ values between 1.2 and 2 using Equation \ref{Equation3} in each GCLF, excluding the counts of magnitude with $<80\%$ completeness (shown by the gray data points in Figure \ref{fig:GCLF}). These best-fit parameters are shown in Table \ref{tab:sigma values}. Using these best-fit parameters, we model the intrinsic GCLF using Equation \ref{Equation3} without the recovery rate ($f(m)$), which are shown by the solid colored lines in Figure \ref{fig:GCLF}. However, as mentioned in \citet{Harris2024p2}, the $m_{0}$ values in Table \ref{tab:sigma values} are brighter than expected due to our methods of deriving the recovery rate. Further analysis of these parameters can be done by finding the completeness of each individual GC candidate using Neural Networks \citep{HarrisNeural}.

\subsubsection{GCLF of Color-Selected GCs}
\label{subsub:GCLF Color Selected}
The GCLF of our color-selected GC candidates (left panel of Figure \ref{fig:GCLF}) follows a lognormal shape, which is the expected shape for luminosity distributions of GCs \citep{Hanes1977, Harris1981}. The reduced $\chi^2$ values, as shown in  Table \ref{tab:sigma values}, gives a best-fit at $\sigma = 1.2$, $N_{col} = 328 \pm 42$, and $m_{0, col} = 30.02 \pm 0.12$. The best-fit turnover magnitude corresponds to $M_{col} = -11.15 \pm 0.18$ mag, given an assumed distance modulus of $m-M = 41.18$ mag. However, this best-fit turnover magnitude is brighter than expected for G165. Given that F200W is closest to the K-band, using the turnover magnitude for nearby galaxies ($M_{V} \simeq -7.5$), the typical color for old stellar populations (V$-$K = 2.8 mag; \citealt{Girardi1998}), and the conversion from Vega mag to AB mag (K[AB] - K[Vega] = 1.8), gives an estimated turnover magnitude for F200W in G165 of $M_{\rm F200W} \approx -7.5 -2.8 + 1.8 = -8.5$ mag ($m_{\rm F200W} \approx 32.7$ mag). This indicates that the $\sigma = 2.0$ fit, which yields a turnover magnitude of $M_{\rm F200W} = -8.13 \pm 0.30$ mag, may be more robust for the color-selected GC sample.

\startlongtable
\begin{deluxetable*}{c|cc|cc|cc|cc}
\tabletypesize{\scriptsize}
\centering
\tablecaption{Best-fit parameters as a function of assumed $\sigma$ \label{tab:sigma values}}
\tablehead{
\colhead{\normalsize $\sigma$} \vline & \colhead{\normalsize $N_{col}$} & \colhead{\normalsize $N_{ang}$} \vline & \colhead{\normalsize $m_{0, col}$} & \colhead{\normalsize $m_{0, ang}$} \vline & \colhead{\normalsize $M_{col}$} & \colhead{\normalsize $M_{ang}$} \vline & \colhead{\normalsize $\chi^{2}_{\nu, \rm col}$} & \colhead{\normalsize $\chi^{2}_{\nu, \rm ang}$}\\
\colhead{[AB Mag]} \vline & \colhead{} & \colhead{} \vline & \multicolumn{2}{c}{[F200W 0.18" mag]}
\vline& 
\multicolumn{2}{c}{[F200W 0.18" mag]}
\vline 
}
\startdata
$1.2$ & $328 \pm 42$ & $1658 \pm 835$ & $30.03 \pm 0.12$ & $31.32 \pm 0.23$ & $-11.15 \pm 0.18$ & $-9.86 \pm 0.27$ & $1.155$ & $1.199$\\
$1.4$ & $520 \pm 81$ & $4809 \pm 3631$ & $30.63 \pm 0.14$ & $32.31 \pm 0.30$ & $-10.55 \pm 0.20$ & $-8.88 \pm 0.33$ & $1.164$ & $1.269$ \\
$1.6$ & $873 \pm 189$ & $16208 \pm 16655$ & $31.33 \pm 0.19$ & $33.46 \pm 0.37$ & $-9.85 \pm 0.23$ & $-7.73 \pm 0.40$ & $1.252$ & $1.336$\\
$1.8$ & $1558 \pm 473$ & $63372 \pm 26215$ & $32.14 \pm 0.23$ & $34.76 \pm 0.28$ & $-9.04 \pm 0.27$ & $-6.42 \pm 0.31$ & $1.364$ & $1.397$\\
$2.0$ & $2920 \pm 1006$ & $99926 \pm 22442$ & $33.05 \pm 0.27$ & $35.63 \pm 0.19$ & $-8.13 \pm 0.30$ & $-5.55 \pm 0.24$ & $1.456$ & $1.302$\\
\enddata
\end{deluxetable*}

\begin{figure*}[ht!]
\centering
\includegraphics[width = 1\textwidth]{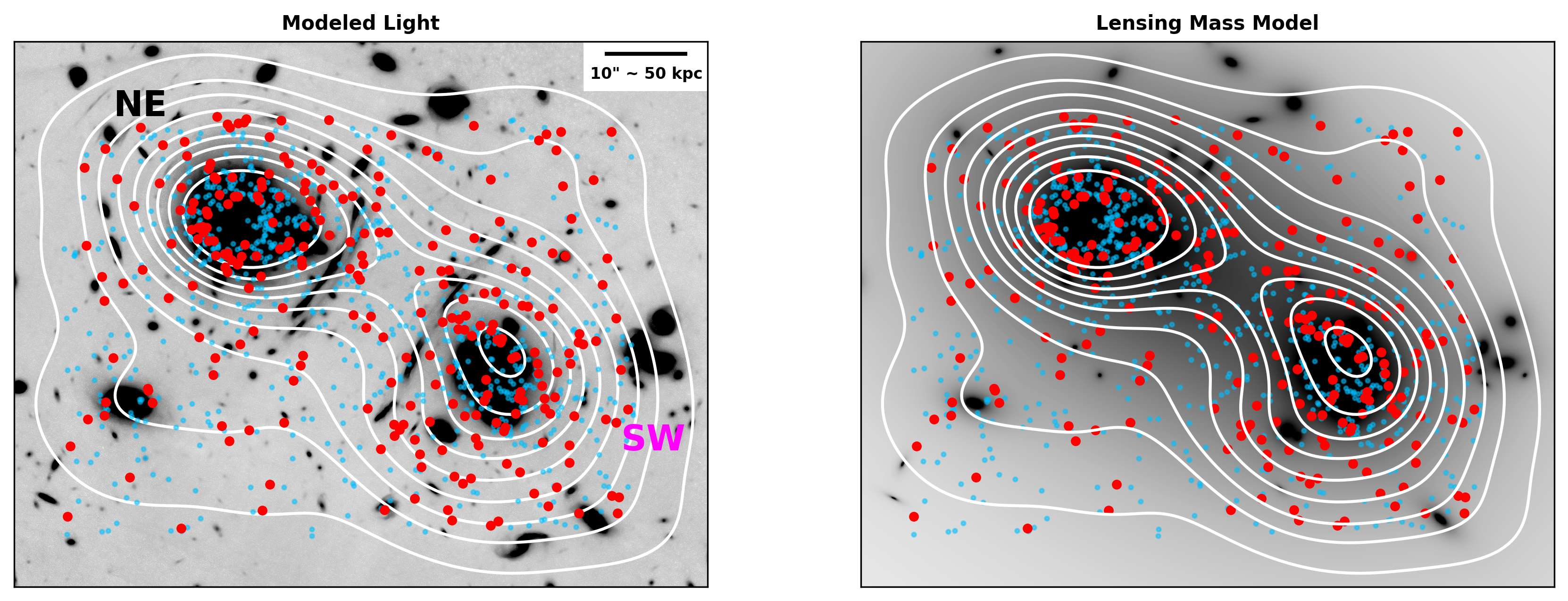}
\caption{
The comparison of our color-selected GC sample spatial distribution with the diffuse light modeled with ProPane (\emph{Left Panel}) and the lensing mass (\emph{Right Panel}), each shown by the respective background images. The red points are the color-selected GCs in our sample, the smaller blue points are all detected point sources (even past the 50$\%$ completeness limit), and the white lines show the contours describing the spatial distribution of the color-selected GC candidates.
\label{fig:ICL Density Map}
}
\end{figure*}

\subsubsection{GCLF of Angular-Selected GC Sample}
\label{subsub:ang}
The GCLF of our GC candidates selected by angular separation (described in Section \ref{subsec:Selection}) is shown in the right panel of Figure \ref{fig:GCLF}. Like the GCLF of our color-selected GC candidates, we fit the GCLF of interloper-subtracted GC candidates using fixed $\sigma$ values from 1.2 to 2.0.  The minimum reduced $\chi^2$ value, as shown in Table \ref{tab:sigma values}, shows a best fit of $\sigma = 1.2$, $N_{ang}= 1658 \pm 835$, and $m_{0, ang} = 31.32 \pm 0.23$ mag ($M_{ang} = -9.86 \pm 0.27$ mag). As with the best-fit parameters for the color-selected GC sample, this best-fit turnover magnitude is also brighter than the expected turnover magnitude for G165, indicating that the best fit may be closer to $\sigma = 1.4$ which yields a turnover magnitude of $M_{ang} = -8.88 \pm 0.33$ mag.

\subsubsection{K-corrections}
K-corrections are necessary to compare rest-frame luminosities across redshifts in galaxies \citep{Oke1968,Pence1976,Kinney1996, Blanton2007}, and are likewise necessary for GC systems at cosmological distances. Using RESCUER \citep{Rescuer}, we obtain a K-correction of K$_{\rm F200W}$ = -0.44 mag, assuming a BASTI isochrone model, metallicity of [M/H] = -1, and GC age $\tau = $ 8 Gyr. Applying this K-correction to the turnover magnitudes at the minimum $\chi^2$ fits gives $M_{ang} = -9.86 +0.44 = -9.42$ mag and $M_{col} = -11.15 + 0.44 = -10.71$ mag. As pointed out in Sections \ref{subsub:GCLF Color Selected} and \ref{subsub:ang}, the expected turnover magnitude for each sample should be closer to $M_{ang} = -8.88 +0.44 = -8.44$ mag and $M_{col} = -8.13 + 0.44 = -7.69$ mag, given the expected K-corrected turnover magnitude for G165 is $M = -8.5 +0.44 = -8.06$. Using the same assumptions of [M/H] and $\tau$, we use Equation 5 of \citet{Harris2024p2} to show that the turnover magnitudes selected by the minimum $\chi^2$ values roughly correspond to a turnover mass estimation between $4-10\times 10^5 M_{\odot}$, which is about 4-5 times higher than the observed value of 1-2 $\times 10^5 M_{\odot}$ for nearby galaxies \citep{Jordan2007, Villegas2010, Harris2014}. The turnover magnitudes closer to expectation for G165 roughly correspond to a turnover mass estimation of $0.6-1.2\times 10^5 M_{\odot}$, closely matching the observed turnover mass in nearby galaxies. However, our estimation of the turnover masses is very uncertain due to difference in redshift between Abell 2744 and G165, and the use of a different filter (F150W) in Abell 2744. Future observations of G165 may bring the observed peak of the GCLF closer to the true turnover point. Once that is the case, constraints of the turnover mass corresponding to the turnover magnitudes of G165 can be more reasonably done by applying the PARSEC CMD3.7 stellar models \citep{Bressan2012, Marigo2013} to the data.

\subsubsection{Predicted Number of GCs}
\label{subsub:numberGCs}
Previous studies \citep{Harris2017, Diego2024} have derived calculations of estimated GC counts of galaxy clusters, given their mass. Specifically, \citet{Diego2024} (see their Sec. 6) found that:
\raggedbottom
\begin{equation}
  N_{GC} \sim 8.3\times10^{-11} \cdot M_{\rm tot}/(M_\odot) 
 \label{Equation4},
\end{equation}

\noindent
where $M_{\rm tot}$ is the total mass of the galaxy cluster and $N_{GC}$ is the expected number of GCs in the galaxy cluster. Using the virial mass of G165, $M_{vir} \simeq 1.3 \times 10^{15} M_{\odot}$ \citep{Frye2019}, we find an expectation of $N_{GC} \approx 1.1\times10^{5}$ GCs in G165. Furthermore, we perform another independent test using the relationship between the specific frequency of the host galaxies and their luminosities \citep{Harris1981} to roughly derive an expected number of GCs. Assuming a specific frequency of $S_{N} = 3.5$ (the average frequency for elliptical galaxies; \citealt{Harris1981, Miller1997}) and using the absolute magnitudes of the four brightest cluster members in G165, found in Table 3 of \citet{Frye2024:aa}, a rough estimate of $N_{GC} \approx 1.1 \times 10^{5}$ GCs is also found. In Table \ref{tab:sigma values}, it is clear that the estimated number of GCs from our GCLF models, $N_{col} \approx 320-2900$ and $N_{ang} \approx 1700-100,000$, are lower than expected, especially for the color-selected GC sample. There are various reasons for the underestimation, including the restrictiveness of our point source and GC selection, limitations of subtracting ICL close to the cluster cores, and insufficient depth of the data. Future deeper observations of G165 will mitigate these limitations, enabling a more robust study of its GC systems in future work.

\subsection{GC Spatial Distribution} \label{subsec:Spatial Comparison}
Observations of the spatial distribution of GCs has historically been used to trace the diffuse light and mass distribution of galaxies and galaxy clusters in the past decades \citep{Tamura2006,Brodie2014,Durrell2014,Napolitano2014,LeeJang2016, Wang2020,Lee2022,Sanchez2022,Martis2024:aa, Kluge2025}. Following similar approaches, we compare the spatial distribution of the color-selected GC sample to the diffuse light and mass distribution of G165. The left panel of Figure \ref{fig:ICL Density Map} shows the comparison of our color-selected GC candidate sample to the modeled light by ProPane, focused on the brightest galaxies in G165. We use Kernel Density Estimation \citep{KernelDensity} in \textsc{seaborn} \citep{Waskom2021} to determine the density of the color-selected GC candidates. By eye, the contours show that there is broad agreement between the spatial distribution of our GC candidate sample and ICL in G165, especially close to the cores of the brightest galaxies \citep{Pascale2022b,Lee2022,Diego2023,   Martis2024:aa}.

In the right panel of Figure \ref{fig:ICL Density Map} we compare the spatial distribution of our color-selected GC candidates to the total mass distribution for G165, given by a recent lens model presented in \citet{Kamieneski2024}. We also find similarities between the lensing mass and spatial distribution of GC candidates, which follows earlier studies of other galaxy clusters using JWST and HST (\citealt{LeeJang2016, Martinez2013, Lee2022, Diego2024, Martis2024:aa}). Likewise, these results compare well with simulations of GC systems \citep{Reina-Campos2022, Reina-Campos2023}, which show the effectiveness of using GC spatial distributions to trace dark matter halos.

\begin{figure}[H]
\hspace*{-0.3cm}
\centering
\includegraphics[width=0.5\textwidth]{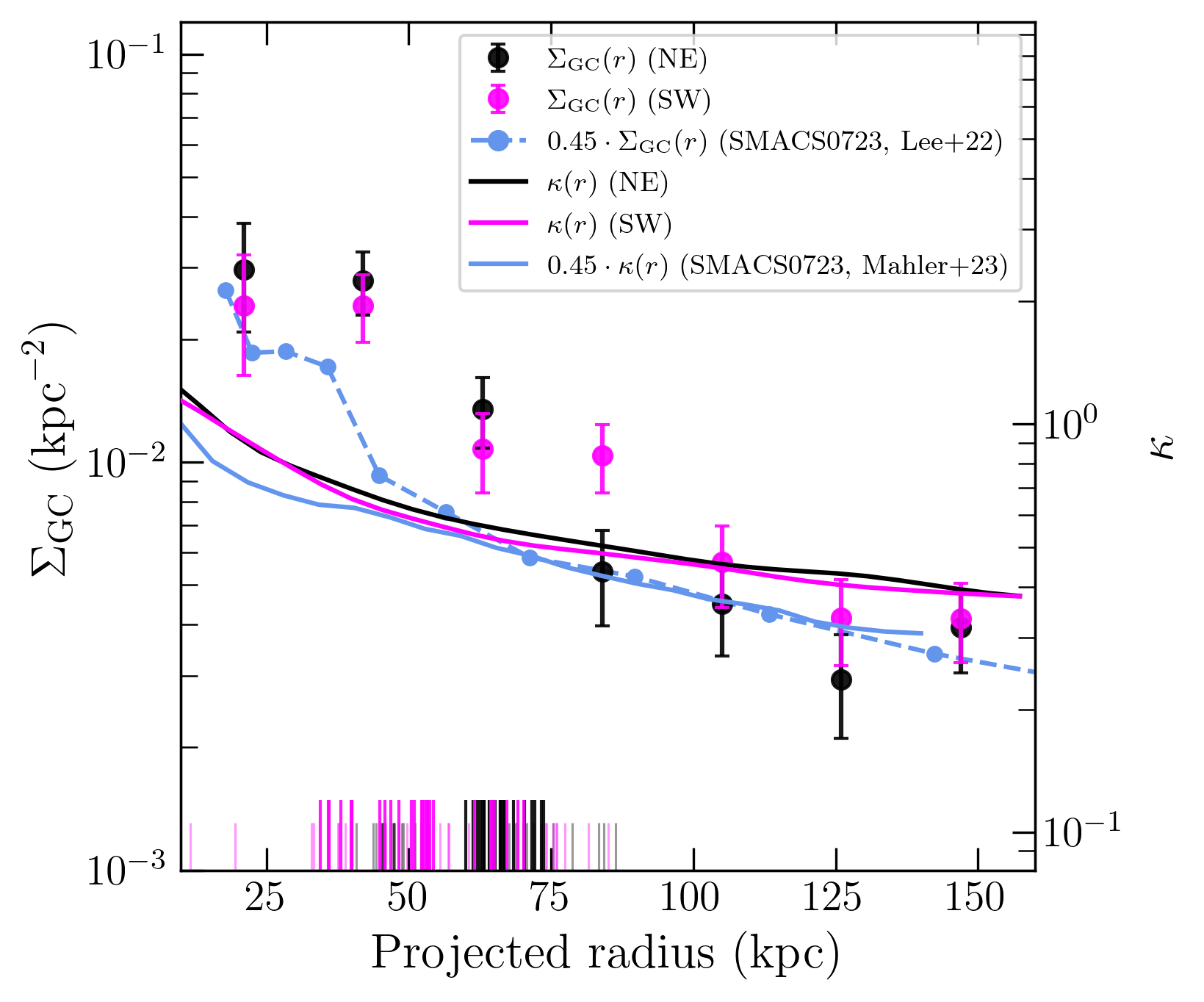}
\caption{Radial number density of color-selected GC candidates to the mass distribution of the NE and SW halos. The magenta and black data points show the number densities, with Poisson uncertainties, of the color-selected GC candidates in the SW and NE halo of G165, respectively. Likewise, the solid magenta and black lines show the convergence of the SW and NE halos from a G165 lensing model \citep{Kamieneski2024}, which correspond to the secondary vertical axis on the right (the limits of which are chosen to approximately align the convergence curves with the GC candidate number densities between $60-120$ kpc). The light blue data points and solid line show the number densities of GC candidates and the convergence in SMACS0723 \citep{Lee2022, Mahler2023},
both rescaled by a factor of 0.45 to account for the average mass ratio of the two individual G165 cores to SMACS0723. The rug plot at the bottom of the figure shows the positions of the arcs in the 21 lensed image families used to constrain the lens model \citep{Frye2024:aa}, colored magenta or black according to the nearest halo component. Taller, thicker lines indicate those arc systems with spectroscopically confirmed redshifts.
\label{fig:radial}
}
\end{figure}

Figure \ref{fig:radial} shows the quantitative comparison between the surface number density $\Sigma_{\rm GC}$ of our color-selected GC candidates and the 
surface mass density, or convergence $\kappa$, inferred from the lens model.
Both are shown as radial profiles
at radially-increasing annuli measured in kpc, centered on the NE and SW cluster cores indicated in Figure \ref{fig:ICL Density Map}. The number densities of both halos (magenta and black data points) are scaled by $\frac{N_{tot}}{N_{sample}}$ to account for the incompleteness of our restrictive selection technique, where $N_{tot}$ is the maximum expected number of GCs in G165 from our GCLF curve fitting ($N_{tot} = 2920$), as indicated by $N_{col}$ in Table \ref{tab:sigma values}, and $N_{sample}$ is the number of color-selected GC candidates ($N_{sample} = 972$). 

The limits of the secondary vertical axis on the right (corresponding to lens model-derived convergence; \citealt{Kamieneski2024}) are chosen to approximately align the curves with the $\Sigma_{\rm GC}$ profiles between $60 - 120$ kpc.
To compare with previous studies, we also include the number densities of GC candidates and convergence of SMACS0723 \citep{Lee2022} using the \citet{Mahler2023} model (constructed with similar methodology as the model from \citealt{Kamieneski2024}). According to these respective lens models, the total mass within projected 400 kpc for SMACS0723 is very similar to G165 ($\approx 3.5\times10^{14}~M_\odot$ for SMACS0723 and $\approx 3.1\times10^{14}~M_\odot$ for G165).For a more direct comparison, in Figure~\ref{fig:radial} we rescale the SMACS0723 GC density and convergence by a factor of 0.45, which is the average mass ratio of the two G165 cores to the total mass of SMACS0723. Even with the limitations near the cores of the brightest cluster members (out to $\sim$ 10 kpc) given our ICL subtraction and limitations of our color selection, we find comparable results to \citet{LeeJang2016, Lee2022, Reina-Campos2022, Diego2023, Diego2024}, in that GC densities decline consistently as a function of cluster-centric radius beyond 60 kpc out to at least 150 kpc, seen in both cluster halos. However, the GC profiles are steeper than their respective convergence profiles, with $\sigma_{GC} \sim \kappa(r)^{3.2 - 3.7}$, with the power-law slopes for the two cores in agreement within uncertainties. This elevated concentration of GC candidates relative to the total cluster mass distribution may be the result of a selection bias in identifying GCs or a real physical effect, but it is difficult to determine at present with the number of candidates that are recovered. 

In Figure \ref{fig:ICL Density Map}, we also show all sources that pass the compactness criteria described in Section \ref{subsec:photometry}, designated by blue points. By eye, it is clear that the point source sample follows the contours of the color-selected GC candidate sample, further indicating that our color-selected sample includes reliable GC candidates, but the color selection leaves out many other GC candidates in the point source sample.

\subsection{Dynamics of G165 using GC spatial distribution}
\label{subsec:G165 Dynamics}
G165 is suspected to be an ongoing merging galaxy cluster in the plane of the sky due to the orientation of radio jets observed in the easternmost cluster members and the single velocity peak of cluster members, indicating that the groups of cluster members have already traversed \citep{Pascale2022a}. The low number of GC candidates between the prominent groups of galaxies in G165, as seen in Figure \ref{fig:ICL Density Map}, suggests that the GCs have not yet been stripped far away from their host galaxies, indicating that the traversing of these two groups of galaxies has only recently occurred \citep{Diego2023, Lee2022, Martis2024:aa}. We caution that it is possible that the small number of GC candidates between the groups of galaxies is driven by poor completeness, given the 50$\%$ completeness limit is much brighter than the expected turnover magnitude for G165, as described in Section \ref{subsec:GCLF}. Further analyses of the halos of G165 cluster members, along with future X-ray observations (M. Donahue et al., in prep), will help further constrain the dynamical state of the galaxy cluster \citep{Pascale2022a}.

\section{Summary and Conclusions} \label{sec:conc}
In this study, we used 3-epoch deep NIRCam JWST observations of the G165 merging galaxy cluster at $z$ = 0.35 to analyze its globular cluster systems. A summary of our findings is as follows:
\begin{itemize}
    \item Using \textsc{ProPane}, we model and subtract the background light (primarily ICL) of G165, enabling us to effectively detect GCs. Using the ICL-subtracted images, we detect sources using \textsc{SExtractor}, select point sources based on their compactness, and use the colors of the point sources in comparison with simple stellar populations at different ages to select the globular cluster candidates. In addition to the color selection, we also identify GC candidates by subtracting background interlopers from the point source sample. A 50$\%$ detection completeness limit of 29.73 mag is reached with the stacked 3-epoch observations.
    \item Using the color and angular separation selected GC candidates, we derive empirical GCLFs to estimate the intrinsic turnover magnitudes estimated from fixed fitting values of dispersion in F200W. Using the minimum $\chi^2$ values from the fits, we predict the best-fit K-corrected turnover magnitudes to be $M_{F200W} = -9.42$ mag for the angular-selected GC sample and $M_{F200W} = -10.71$ mag for the color-selected GC sample. The expected turnover magnitude for G165 should be closer to $M_{\rm F200W} \approx -8.5$ mag, fainter than our modeled values. This highlights the need for deeper imaging to constrain the GCLF turnover in G165.
    \item We compare the spatial distribution of the color-selected GCs candidates with the ICL modeled with ProPane and the lensing mass of G165 using a recently derived lens model. We find there is qualitative correlation between these three, especially in the vicinity of the galaxy cores. Further quantitative evidence of correlation between the lensing mass and color-selected candidates is shown by the radial number densities of color-selected GC candidates to the convergence of G165, which also match the respective radial profiles derived for the SMACS0723 galaxy cluster\citep{Lee2022}.

\end{itemize}

Future analyses of globular clusters can also be done in many other galaxy clusters observed by JWST. Over 30 galaxy clusters have now been observed with JWST, spanning a redshift of 0.25 to $>$1 (e.g. \citealt{Windhorst2023:aa}, \citealt{Willott2023}, \citealt{slice2025}). Further studies of GCs in these fields will advance our knowledge of the GCLF and the connection between the distribution of GCs and mass at higher redshifts.
\\
\begin{acknowledgments} 
The authors would like to sincerely thank Katherine Whitaker, Steven Willner, Massimo Ricotti, and Cheng Cheng for the helpful discussions and suggestions that helped to improve the manuscript. This work is based on observations made with the NASA/ESA/CSA James Webb Space Telescope. The data were obtained from the Mikulski Archive for Space Telescopes at the Space Telescope Science Institute, which is operated by the Association of Universities for Research in Astronomy, Inc., under NASA contract NAS 5-03127 for JWST. These observations are associated with JWST programs \#~1176 (PEARLS;PI: R. Windhorst) and \#~4446 (PI: B. Frye). The specific observations analyzed can be accessed via \dataset[doi: 10.17909/dc5a-ta61]{https://doi.org/10.17909/dc5a-ta61.} TRH acknowledges partial support from the Arizona NASA Space Grant Consortium, Cooperative Agreement 80NSSC20M0041. R.A.W., S.H.C., and R.A.J. acknowledge support from NASA JWST Interdisciplinary Scientist grants NAG5-12460, NNX14AN10G and 80NSSC18K0200 from GSFC. WEH acknowledges support from the Natural Sciences and Engineering Research Council of Canada.

\end{acknowledgments}
\facilities{JWST (NIRCam, NIRSpec)}

\software{{\textsc{astropy}} \citep{astropy1,astropy2,astropy3}, {\textsc{eazy}} \citep{BrammerEAZY}, {\textsc{glue}} \citep{glue1, glue2}, {Ned Wright's Cosmology Calculator} \citep{Wright2006}, {\textsc{photutils}} \citep{Bradley2016}, {\textsc{ProPane}} \citep{Robotham2024:aa}}

\appendix

\section{ICL Subtraction Methods} \label{app:Isophote}
\begin{figure}[H]
\centering
\includegraphics[width=1\textwidth]{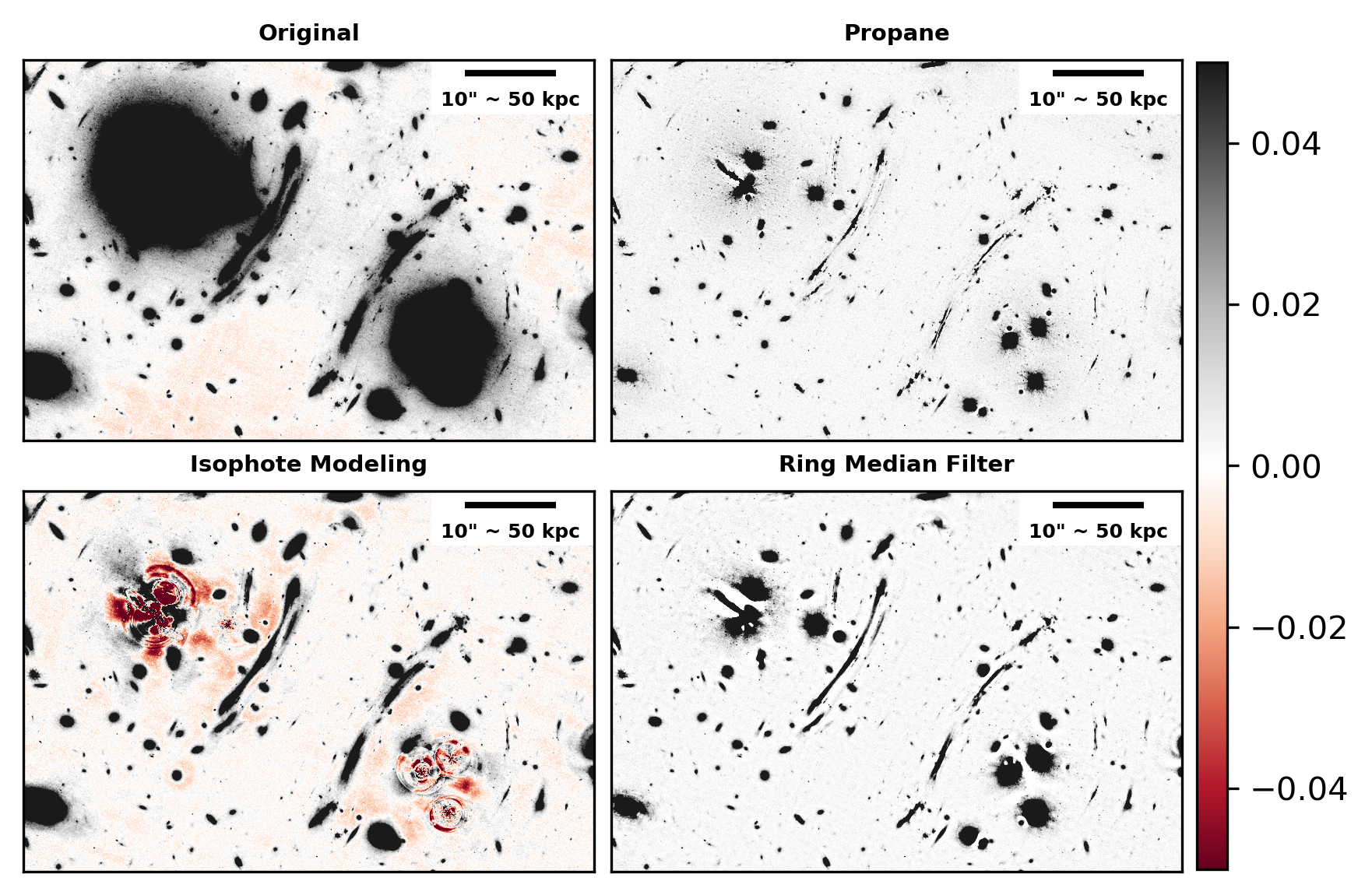}
\caption{
Comparison of residuals found in F200W using different ICL subtraction methods. \emph{Top Left Panel:} Original image without any ICL or diffuse light subtraction. Negative values are an artifact of sky subtraction in the original data reduction. \emph{Top Right Panel:} ICL subtraction using \textsc{ProPane}, the primary method used in this study. \emph{Bottom Left Panel:} ICL subtraction using \textsc{Photutils}' Isophote Modeling. \emph{Bottom Right Panel:} ICL subtraction using ring median filtering ($r_{in}$ = 18, $r_{out}$ = 20), as described in \citet{Lee2022}. See also Fig. 3 in \citet{Pascale2022a} for residuals when using \textsc{galfit}.
\label{appfig:ICL Methods}
}
\end{figure}

In Figure \ref{appfig:ICL Methods}, we show the comparison of different ICL subtraction methods, including Isophote Modeling, \textsc{ProPane}, and ring median filtering. Isophote modeling fits each galaxy using elliptical isophotes, which we applied to the 7 brightest galaxies in G165. We find that Isophote Modeling has the most oversubtraction, but performed best in removing isolated, elliptical galaxies. Likewise, we find that ring median filtering using ring sizes of $r_{in}$ = 18 pixels and $r_{out}$ = 20 pixels tends to subtract too much light between compact groups of galaxies, which is a significant feature of G165. Although ProPane only yields positive values in the background light subtraction, it seems to perform better than the other two methods in ensuring a uniform background level for point-source identification.

\section{GC Color Selection Bounds $\&$ Photometric Redshift Object Properties} \label{app:Bounds}
\begin{figure*}[htb!]
\centering
\includegraphics[width=0.5\textwidth]{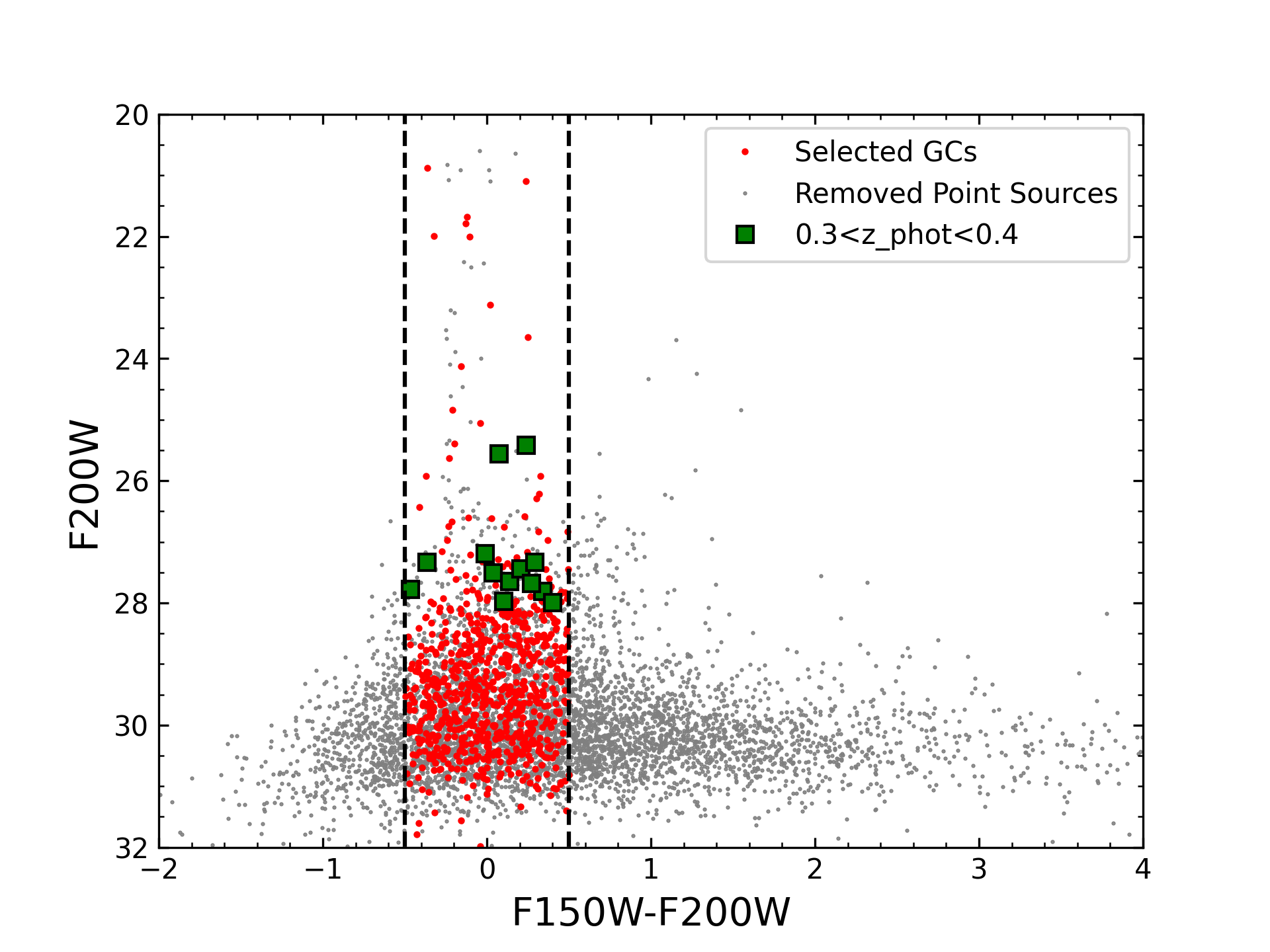}\hfill \includegraphics[width=0.5\textwidth]{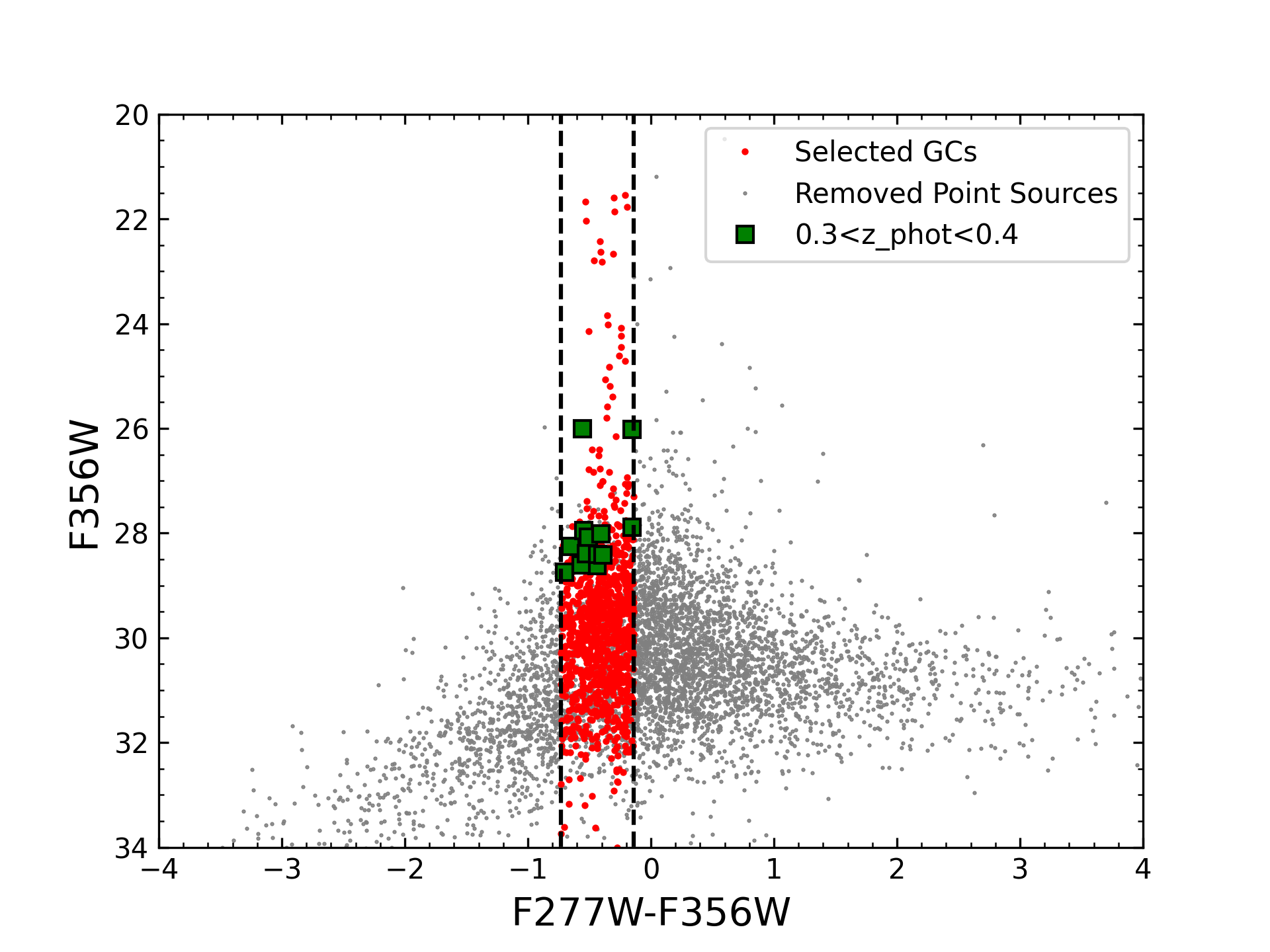}
\caption{
Color magnitude diagrams (Kron apertures) used to constrain the bounds of the GC color selection. Red points show the selected GC candidates, gray points show the removed point sources, and green squares are point sources with photometric redshifts between 0.3 and 0.4. \textit{Left Panel:} Color magnitude diagram using F150W and F200W. \textit{Right Panel:} Color magnitude diagram using F277W and F356W.
 \label{fig:appCMD}
}
\end{figure*}

In Figure \ref{fig:appCMD}, we show the color magnitude diagrams used to motivate the bounds for our color selection of GC candidates in Figure \ref{fig:GC Selection}. The bounds were determined to bracket the well-fit photometric redshift objects in the CMDs of F150W$-$F200W and F277W$-$F356W. The two outliers at F277W-F356W $\sim$ -0.2 are left out due to the higher redshifts at this value, given by the stellar population tracks in Figure \ref{fig:GC Selection}. Along with these color magnitude diagrams, we include basic information about the 13 objects selected by their photometric redshifts (green squares in Figure \ref{fig:appCMD}) in Table \ref{tab:photz objects}, including their coordinates, apparent magnitude in F200W, and photometric redshift with their errors found using \textsc{EAZY}.

\startlongtable
\begin{deluxetable*}{cccc}
\tabletypesize{\scriptsize}
\centering
\tablecaption{Properties of objects with well-fit photometric redshifts\label{tab:photz objects}}
\tablehead{
\colhead{\normalsize R.A. (J2000)} & \colhead{\normalsize Decl. (J2000)}  & \colhead{\normalsize $m_{\rm F200W}$} & \colhead{\normalsize $z_{\rm phot}$} \\
\colhead{[Deg]} & \colhead{[Deg]} & \colhead{[0.18" mag]} & \colhead{}
}
\startdata
$171.821095$ & $42.464749$ & $27.86 \pm 0.02$ & $0.31^{+0.07}_{-0.05}$\\
$171.818577$ & $42.466098$ & $26.02\pm 0.003$ & $0.39^{+0.02}_{-0.02}$\\
$171.818294$ & $42.467883$ & $28.26\pm 0.02$ & $0.35^{+0.09}_{-0.05}$\\
$171.812878$ & $42.469948$ & $27.97\pm 0.02$ & $0.36^{+0.03}_{-0.04}$\\
$171.747597$ & $42.470325$ & $27.46\pm 0.01$ & $0.35^{+0.09}_{-0.01}$\\
$171.772910$ & $42.470392$ & $27.44\pm 0.01$ & $0.38^{+0.03}_{-0.04}$\\
$171.833823$ & $42.475214$ & $27.88\pm 0.02$ & $0.35^{+0.04}_{-0.07}$\\
$171.810126$ & $42.475937$ & $27.47\pm 0.01$ & $0.39^{+0.03}_{-0.03}$\\
$171.813755$ & $42.476448$ & $27.97\pm 0.02$ & $0.31^{+0.03}_{-0.04}$\\
$171.817086$ & $42.479555$ & $25.87\pm 0.002$ & $0.38^{+0.02}_{-0.02}$\\
$171.752101$ & $42.480622$ & $27.51\pm 0.01$ & $0.31^{+0.03}_{-0.02}$\\
$171.754912$ & $42.480627$ & $28.22\pm 0.02$ & $0.32^{+0.04}_{-0.04}$\\
$171.797209$ & $42.494500$ & $27.84\pm 0.01$ & $0.38^{+0.04}_{-0.04}$\\
\enddata
\end{deluxetable*}
\bibliography{References}

\end{document}